\documentclass[epj]{webofc}
\usepackage[utf8]{inputenc}
\usepackage[varg]{txfonts}   
\usepackage{booktabs}
\usepackage{xcolor}
\definecolor{darkred}{rgb}{0.4,0.0,0.0}
\definecolor{darkgreen}{rgb}{0.0,0.4,0.0}
\definecolor{darkblue}{rgb}{0.0,0.0,0.4}
\usepackage[bookmarks,linktocpage,colorlinks,
    linkcolor = darkred,
    urlcolor  = darkblue,
    citecolor = darkgreen]{hyperref}
\usepackage{subfigure}
\wocname{EPJ Web of Conferences}
\woctitle{Lattice2017}
\begin{document}
%
\selectlanguage{english}
\title{%
  Can axial $U(1)$ anomaly disappear at high temperature?
}
\author{%
  \firstname{Hidenori} \lastname{Fukaya}\inst{1}\fnsep
  \thanks{Speaker, \email{hfukaya@het.phys.sci.osaka-u.ac.jp}}
\firstname{} \lastname{for JLQCD collaboration}
}
\institute{%
Department of Physics, Osaka University, Toyonaka 560-0043, Japan
}
\abstract{%
  In our recent study of two-flavor lattice QCD using chiral fermions,
  we find strong suppression of axial $U(1)$ anomaly above the critical temperature of
  chiral phase transition. Our simulation data also indicate suppression of
  topological susceptibility. In this talk, 
  we present both of our theoretical and numerical evidence for disappearance
  of axial $U(1)$ anomaly, emphasizing the importance of controlling lattice
  chiral symmetry violation, which
  is enhanced at high temperature.
}
\maketitle
\section{Introduction}\label{intro}

``Can axial $U(1)$ anomaly disappear at high temperature?''
To this question the standard answer would be ``No.''
The reason is that unlike spontaneous
breaking of symmetry, anomaly is a symmetry breaking
at the cut-off scale where the theory is defined,
and the anomalous Ward-Takahashi identity (WTI) with any operator $O(x')$,
\begin{eqnarray}
  \label{eq:AWTI}
  \langle \partial_\mu J^\mu_5 (x) O(x')\rangle_{f}
  -\langle \delta_A O(x)\rangle_{f}\delta(x-x')
  &=& 
\frac{N_f}{32\pi^2} \epsilon_{\mu\nu\rho\sigma}F^{\mu\nu}F^{\rho\sigma}(x) \langle O(x')\rangle_{f}, 
\end{eqnarray}
holds at any energy scale, and for any gluon background, and so on.
Here the flavor singlet axial $U(1)$ (we denote $U(1)_A$ in the following)
transformation is denoted by $\delta_A$,
to which the associated axial current is given by $J^\mu_5 (x)$.
$N_f$ is the number of flavors and $F^{\mu\nu}$ is the field strength of the gluon fields.

The identity (\ref{eq:AWTI}) can not be a complete answer, since
only the fermion part of the QCD path integral is performed,
denoted by $\langle \cdots \rangle_{f}$.
The real question is whether the identity persists to be non-zero
even after the gluon integral $\langle\cdots\rangle_{g}$:
\begin{eqnarray}
  \label{eq:WTIgluons}
  \left\langle\langle \partial_\mu J^\mu_5 (x) O(x')\rangle_{f}
  -\langle \delta_A O(x)\rangle_{f}\delta(x-x')\right\rangle_{g}
  &=& 
\left\langle\frac{N_f}{32\pi^2} \epsilon_{\mu\nu\rho\sigma}F^{\mu\nu}F^{\rho\sigma}(x) \langle O(x')\rangle_{f}\right\rangle_{g} \neq 0 ? 
\end{eqnarray}
In fact, as we will see below for some non-trivial operators $O(x')$,
the right-hand-side(R.H.S.) of (\ref{eq:WTIgluons})
must be zero when the $SU(2)_L\times SU(2)_R$ part of
the chiral symmetry is restored at high temperature.
In this talk, 
we will show that this question is non-trivial
and can only be answered by simulating lattice QCD,
with a careful treatment of the chiral symmetry.

Vanishing of anomaly itself is not surprising as we know an example,
the conformal anomaly which can disappear at the fixed point
of the renormalization group flow.
We should, however, mention that we do not expect 
any $U(1)_A$ symmetric effective quantum field theory to appear at low-energy,
although in the conformal symmetry case, such an effective theory
may appear after taking the so-called scaling limit\footnote{
  We thank M.~L\"uscher for pointing out this difference.
  }.
In fact, we cannot exclude $U(1)_A$ breaking finite volume effects\footnote{
  For this reason, the title of this manuscript is changed from our original
  talk ``Is axial $U(1)$ anomalous at high temperature?''.
}.
Therefore, the $U(1)_A$ symmetric observables, if they exist,
are limited to those surviving the thermodynamical limit,
which is still non-trivial as they could be strong enough to change the phase structure of QCD.

Many previous simulations before 2012 \cite{Bernard:1996iz, Chandrasekharan:1998yx, Ohno:2011yr}
have reported sizable $U(1)_A$ symmetry breaking above critical temperature.
Recent works, on the other hand, show that the lattice artifacts
near the chiral limit is large, and $U(1)_A$ symmetry breaking
in the continuum limit is
smaller than estimated in the literature.
Some of them including us
\cite{Cossu:2013uua, Chiu:2013wwa, Cossu:2015kfa, Brandt:2016daq, Tomiya:2016jwr, Ishikawa:2017nwl, Rohrhofer:2017grg}
concluded that $U(1)_A$ anomaly
is consistent with zero above the critical temperature,
which is however not supported by others
\cite{Bazavov:2012qja, Buchoff:2013nra, Dick:2015twa, Sharma:2016cmz}.
In this talk, we will show that a very precise control of the chiral symmetry,
both in sea and valence quark sectors,
is essential for this problem.

\section{Why and how $U(1)_A$ anomaly can disappear}\label{theory}

Let us begin with integrating (\ref{eq:WTIgluons}) over $x$.
Here we assume that the quark field satisfies the periodic boundary conditions
in the spatial directions, while it is anti-periodic in the temporal direction.
Noting $J_5^\mu$ is periodic, the integration over $x$ leads to
\begin{eqnarray}
  \label{eq:WTIintegrated}
  \left\langle \langle \delta_A O(x')\rangle_{f}\right\rangle_{g}
  &=& 
 - N_f \left\langle Q \langle O(x')\rangle_{f}\right\rangle_{g},
\end{eqnarray}
where $Q$ denotes the topological charge of the gluon fields.
Because of the parity symmetry, $O(x')$ must be parity odd
in order to have a non-trivial expectation value,
and it is natural to assume that its fermionic integral
is expanded in the odd powers of $Q$:
\begin{eqnarray}
  - \left\langle Q \langle O(x')\rangle_{f}\right\rangle_{g} =
  \left\langle Q\left(A \frac{Q}{V}+B\frac{Q^3}{V^2}+\cdots\right)\right\rangle_g
  = A \chi_t + B c_4+\cdots,
\end{eqnarray}
where $\chi_t =\langle Q^2\rangle_g/V$ is the topological susceptibility,
and $c_4 = (\langle Q^4\rangle_g-3\langle Q^2\rangle_g^2)/V$ is the 4-th cumulant, and so on.
The above equations indicate that there are two ways of examining
the $U(1)_A$ anomaly at high temperature: 1) to directly examine the L.H.S. of
(\ref{eq:WTIintegrated}) and 2) to measure $\chi_t$, $c_4$ {\it etc.}
In this work, we try both in our lattice QCD simulations.

Let us examine a simple example, taking $O(x')$ to be a pseudoscalar singlet operator
$\bar{\Psi}i\gamma_5\Psi$. Then the above integrated anomalous WTI gives
\begin{eqnarray}
  \label{eq:scalar}
  \left\langle \langle \bar{\Psi}\Psi \rangle_{f}\right\rangle_{g}
  &=& 
  N_f \left\langle Q \langle \bar{\Psi}i\gamma_5\Psi\rangle_{f}\right\rangle_{g} = \lim_{m\to 0}\frac{N_f^2 \chi_t}{m},
\end{eqnarray}
where $m$ is the quark mass.
The second equation is a consequence of the index theorem.
At zero temperature, the chiral perturbation theory predicts $\chi_t = m \Sigma/N_f $,
which is consistent with $\langle \bar{\Psi}\Psi \rangle= N_f\Sigma$ in the chiral limit.
Note that we have started with the anomalous WTI for the $U(1)_A$ symmetry but
ended up with the chiral condensate, a probe of
spontaneous symmetry breaking (SSB) of $SU(N_f)_L\times SU(N_f)_R$ symmetry.
In fact, these SSB of $SU(N_f)_L\times SU(N_f)_R$ and $U(1)_A$ anomaly  are
tied together for
the quark bi-linear operators:
there is no operator whose expectation value can break $U(1)_A$ without breaking $SU(N_f)_L\times SU(N_f)_R$, and vice versa.
This simple fact indicates that the $U(1)_A$ anomaly and the SSB of $SU(N_f)_L\times SU(N_f)_R$ 
are not independent, but in fact tightly connected.
Another important indication of (\ref{eq:scalar}) is that
in the $SU(N_f)_L\times SU(N_f)_R$ symmetric phase,
$\chi_t$ must vanish as $m^2$ or faster towards the chiral limit.

This simple argument only applies to a single local operator at $x'$.
For non-local or multi-point operators with higher dimensions,
the structure is more complicated.
Not only the zero modes, but also non-zero eigenvalues and
their eigenfunctions are need to examine the symmetry among them. 
There is still a set of operators, which is easy to handle,
as expressed by the eigenvalues only
\cite{Cohen:1996ng, Cohen:1997hz},
\begin{eqnarray}
  \rho(\lambda) = \lim_{V\to \infty}\frac{1}{V}\sum_i \left\langle\langle \delta(\lambda -\lambda_i))\right\rangle_f\rangle_g,
\end{eqnarray}
where $\lambda_i$ denotes the $i$-th eigenvalue of the Dirac operator.
In our work \cite{Aoki:2012yj}, we have investigated the possible
connections between the SSB of $SU(N_f)_L\times SU(N_f)_R$
and the $U(1)_A$ anomaly
through the eigenvalue density of overlap lattice Dirac operator \cite{Neuberger:1997fp,Neuberger:1998my}.
Our work corresponds to a generalization of the Banks-Casher relation \cite{Banks:1979yr}:
\begin{eqnarray}
  -\lim_{V\to \infty}\frac{1}{V}\left\langle \langle \delta_A P^0 \rangle_{f}\right\rangle_{g}
  = \lim_{V\to \infty}\frac{1}{V}\left\langle \langle S^0 \rangle_{f}\right\rangle_{g} = N_f \pi  \rho(0),
\end{eqnarray}
where
\begin{eqnarray}
  P^0 = \int d^4 x\; \bar{\Psi}i\gamma_5\Psi(x),\;\;\;S^0 = \int d^4 x\; \bar{\Psi}\Psi(x).
\end{eqnarray}
In the following, we only consider the case with $N_f=2$.

The first example of such non-trivial operators is
the so-called $U(1)_A$ susceptibility, obtained from
\begin{eqnarray}
  \label{eq:Delta_def}
  \Delta_{\pi-\delta} = \lim_{V\to \infty}\frac{1}{V}\left\langle \langle \delta_A (S^a P^a) \rangle_{f}\right\rangle_{g}
  &=& \lim_{V\to \infty}\frac{1}{V}\left\langle \langle  P^aP^a - S^a S^a \rangle_{f}\right\rangle_{g} 
  = \int_0^\infty d\lambda\;\rho(\lambda)\frac{2m^2}{(\lambda^2+m^2)^2},
\end{eqnarray}
where $P^a, S^a$ are (the $a$-th component of)
the iso-triplet pseudoscalar and scalar operators integrated over space-time,
\begin{eqnarray}
  P^a = \int d^4 x\; \bar{\Psi}i\gamma_5\tau^a\Psi(x),\;\;\;S^a = \int d^4 x\; \bar{\Psi}\tau^a\Psi(x).
\end{eqnarray}
This $U(1)_A$ susceptibility is related to the axial  $SU(2)$ rotation (denoted by $\delta_{SU(2)}$)
of an operator product $S^a P^0$:
\begin{eqnarray}
  \label{eq:N2SU2rot}
  \lim_{V\to \infty}\frac{1}{V}\delta_{SU(2)} \left\langle \langle S^aP^0\rangle_{f}\right\rangle_{g}
  &=& \lim_{V\to \infty}\frac{1}{V} \left\langle \langle P^0P^0-S^aS^a\rangle_{f}\right\rangle_{g}
  \nonumber\\
  &=&
  -\frac{1}{m^2}\chi_t
  + \int_0^\infty d\lambda\;\rho(\lambda)\frac{2m^2}{(\lambda^2+m^2)^2},
\end{eqnarray}
which must vanish in the $m\to 0$ limit when the $SU(N_f)_L\times SU(N_f)_R$ symmetry is restored.
Note that the second term of (\ref{eq:N2SU2rot}) is equivalent to $\Delta_{\pi-\delta}$,
and therefore, $\Delta_{\pi-\delta}$ must vanish unless it coincides with $\chi_t/m^2$.

Similar analysis was done for various operators of the form
$O^{n_1,n_2,n_3,n_4}=(P^a)^{n_1}(S^a)^{n_2}(P^0)^{n_3}(S^0)^{n_4}$
and our result implies
\begin{eqnarray}
  \lim_{m\to 0}\frac{\chi_t}{m^{2N}} =0,\;\;\;\mbox{for any $N$},
\end{eqnarray}
which is equivalent to
\begin{eqnarray}
  \label{eq:mcritical}
  \chi_t = 0\;\;\;\mbox{for $m< {}^\exists m_{cr}$}.
\end{eqnarray}

The examples given above cover only a limited number of operators,
and therefore, far from a proof of vanishing $U(1)_A$ anomaly.
But they demonstrate that SSB of the $SU(N_f)_L\times SU(N_f)_R$ symmetry
and the $U(1)_A$ anomaly are 
tightly connected with each other.
Note here that our argument may be affected by finite volume corrections.
We are not claiming the existence of the low-energy effective theory
manifestly having the $U(1)_A$ ``symmetry'', 
but showing the absence of the $U(1)_A$ anomaly
in the thermodynamical quantities in the large volume limit.
For other theoretical discussions, we refer the readers to 
Refs.\cite{Vicari:2007ma, Nakayama:2014sba,Sato:2014axa,Kanazawa:2015xna,Nakayama:2016jhq}.



\section{Lattice QCD at high temperature with chiral fermions}

In our two-flavor QCD simulations, we employ
the tree-level improved Symanzik gauge action 
and the M\"obius domain-wall fermion action \cite{Brower:2012vk},
which is numerically less expensive than
the overlap fermion and allows topology tunneling at the cost of
violating the Ginsparg-Wilson relation \cite{Ginsparg:1981bj} at some small amount.
We apply the stout smearing \cite{Morningstar:2003gk} three times 
on the gauge links before computing the Dirac operators.

We set the temporal direction $L_t=8,10,12$
at different lattice spacings $a=0.07$--0.1 fm,
which covers the region of temperature $T=190$--330 MeV,
where the critical temperature
is estimated to be around $180$ MeV.
In order to control the systematics due to finite volume, we simulate
three different lattices with the size $L=16,32$ and 48 (2--4 fm).

\begin{figure}[thb] 
  \centering
  \includegraphics[width=10cm]{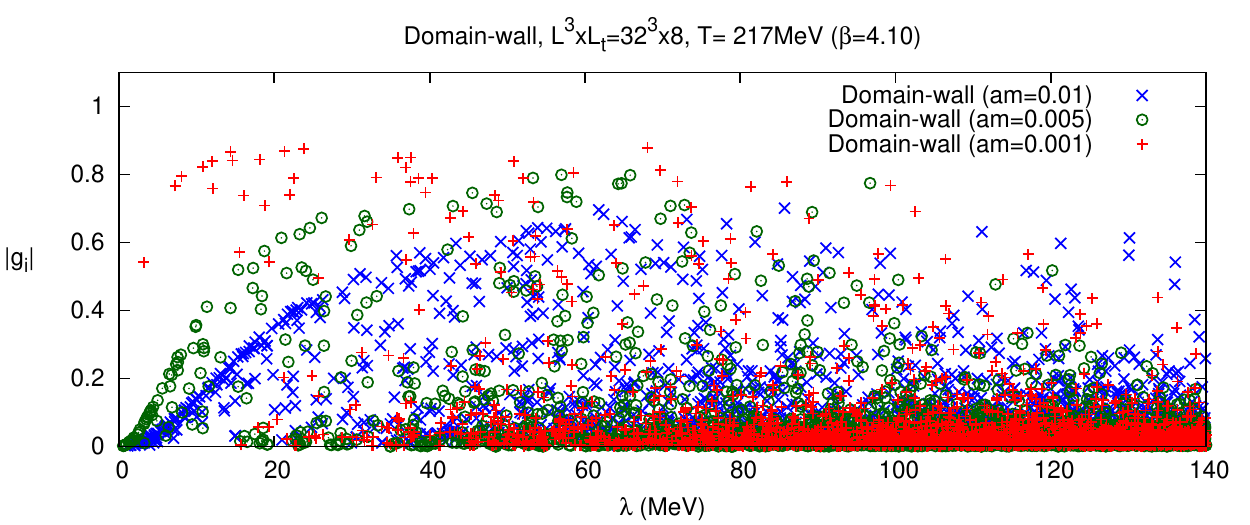}
  \caption{Violation of the Ginsparg-Wilson relation $g_i$ as measured for individual eigenmodes.
    Those for $\beta=4.10$ and  $L^3\times L_t=32^3\times 8$ are shown \cite{Tomiya:2016jwr}.}
  \label{fig:gi}
\end{figure}

In this set-up, the residual mass of the M\"obius domain-wall fermion
is estimated as $1$ MeV.
However, if we look more carefully at individual eigenmodes,
it turns out that there are exceptionally
bad eigenmodes, which strongly violates the Ginsparg-Wilson relation,
as shown in Fig.~\ref{fig:gi}. We plot
\begin{align}
g_i &\equiv
\frac{
\psi_i^\dagger \gamma_5(D_{\rm DW}^{4D}\gamma_5 + \gamma_5D_{\rm DW}^{4D}-2a D_{\rm DW}^{4D}\gamma_5 D_{\rm DW}^{4D})\psi_i}
{\lambda_i^{(m)}}
\left[
\frac{(1-am)^2}{2(1+am)}
\right]
\label{eq:def_gi},
\end{align}
where $\psi_i$ and $\lambda_i^{(m)}$ denote the $i$-th eigenmode and eigenvalue, respectively,
of the massive four-dimensional effective Dirac operator
$\gamma_5((1-m)D_{\rm DW}^{4D}+m)$.
Unfortunately, our target observables,  the $U(1)_A$ susceptibility and topological susceptibility,
are so sensitive to these bad modes that the signals are dominated by the lattice artifacts.
Similar enhancement of the lattice artifact is also reported by two other speakers
\cite{Bonati:2017nhe, Kovacs:lat17}.

Instead, we adopt the overlap fermion
for the $U(1)_A$ related observables, using the overlap/Domain-wall reweighting.
The overlap fermion operator is constructed with exactly treated low-lying eigenmodes of
the kernel operator $H_M$ of the M\"obius domain-wall operator\footnote{
  The kernel is different from the original definition by Neuberger \cite{Neuberger:1997fp, Neuberger:1998my}.
}.
Since the difference between $D_\text{DW}^\text{4D}$ and
$D_\text{ov}$ appears only in the treatment of the low modes of $H_M$,
we expect a good overlap in their relevant configuration spaces,
and a mild fluctuation of the reweighting factor between them.
This is indeed the case for the lattice spacing $a<0.1$ fm.
In order to accumulate sufficient amount of data samples
after the reweighting,
we carry out the run of 
$20000-30000$ molecular dynamics (MD) time,
which also allows the reweighting of the determinant
to that with different sea quark masses.
With these runs, we observe many
topology tunnelings and the auto-correlation time is observed as $O(100)$.

We also find that the use of the overlap fermion only in the valence sector 
(as adopted in \cite{Dick:2015twa, Sharma:2016cmz})
suffers from a significant partially quenching artifacts \cite{Tomiya:2016jwr}.
Therefore, we do not use the overlap fermions without reweighting in the following analysis.


\section{$U(1)_A$ susceptibility}

In this section, we directly investigate 
the $U(1)_A$ anomaly at high temperature by computing the susceptibility $\Delta_{\pi-\delta}$ in (\ref{eq:Delta_def}),
which is obtained from the two-point correlators of the iso-triplet scalar and pseudoscalar channels.
These correlators are related by the $U(1)_A$ symmetry and their difference must vanish
when the symmetry is recovered. The use of the iso-triplet channels has a practical advantage of not including disconnected diagrams,
which are numerically demanding.

We confirm that $g_i$ for the overlap Dirac eigenmodes 
are negligible, and the low-lying modes below $\lambda\sim 0.1/a$ \
are good enough to saturate the sum to obtain $\Delta_{\pi-\delta}$.
Therefore, we can use these low-modes with source points averaged over the whole lattice,
together with the OV/DW reweighting to estimate 
the $U(1)_A$ susceptibility (let us denote it as $\Delta_{\pi-\delta}^{\rm ov}$).

Taking the advantage of good chirality, 
we can subtract the effect of the chiral zero-mode effects:
\begin{align}
\label{eq:Deltabar}
\bar{\Delta}_{\pi-\delta}^{\rm ov} \equiv \Delta_{\pi-\delta}^{\rm ov} - \frac{2N_0}{Vm^2}.
\end{align}
The expectation value of $N_0^2$ is expected to be an $O(V)$ quantity, 
as shown in \cite{Aoki:2012yj},
so that these chiral zero-mode's effects do not survive in the large volume limit,
as $N_0/V$ vanishes as $O(1/\sqrt V)$.
We numerically confirm  the monotonically decreasing volume scaling of 
$\langle N_0/V\rangle$.
Therefore, $\bar{\Delta}_{\pi-\delta}^{\rm ov}$ and $\Delta_{\pi-\delta}^{\rm ov}$ has the same thermodynamical limit.

\begin{figure*}[tbp]
    \centering
    \includegraphics[width=8.0cm]{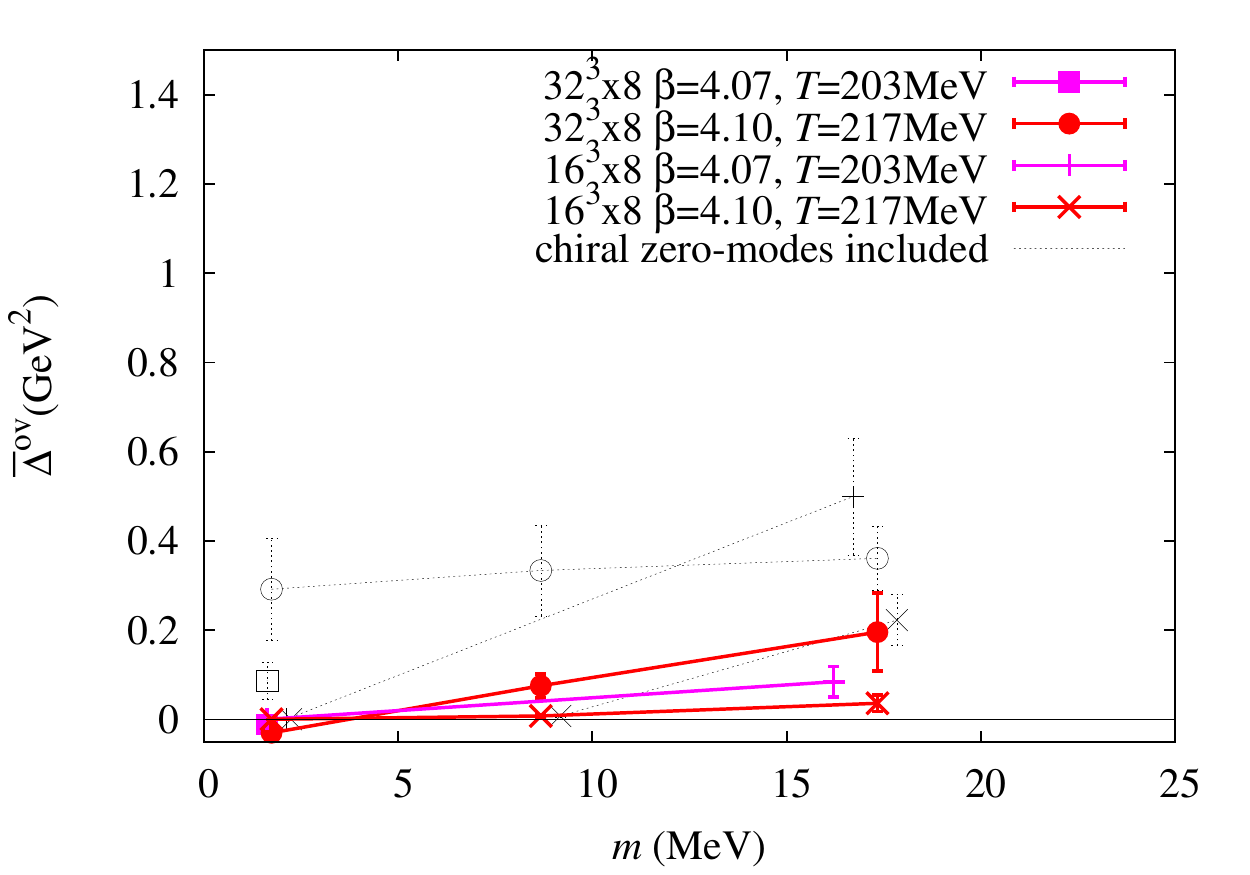}
    \includegraphics[width=8.0cm]{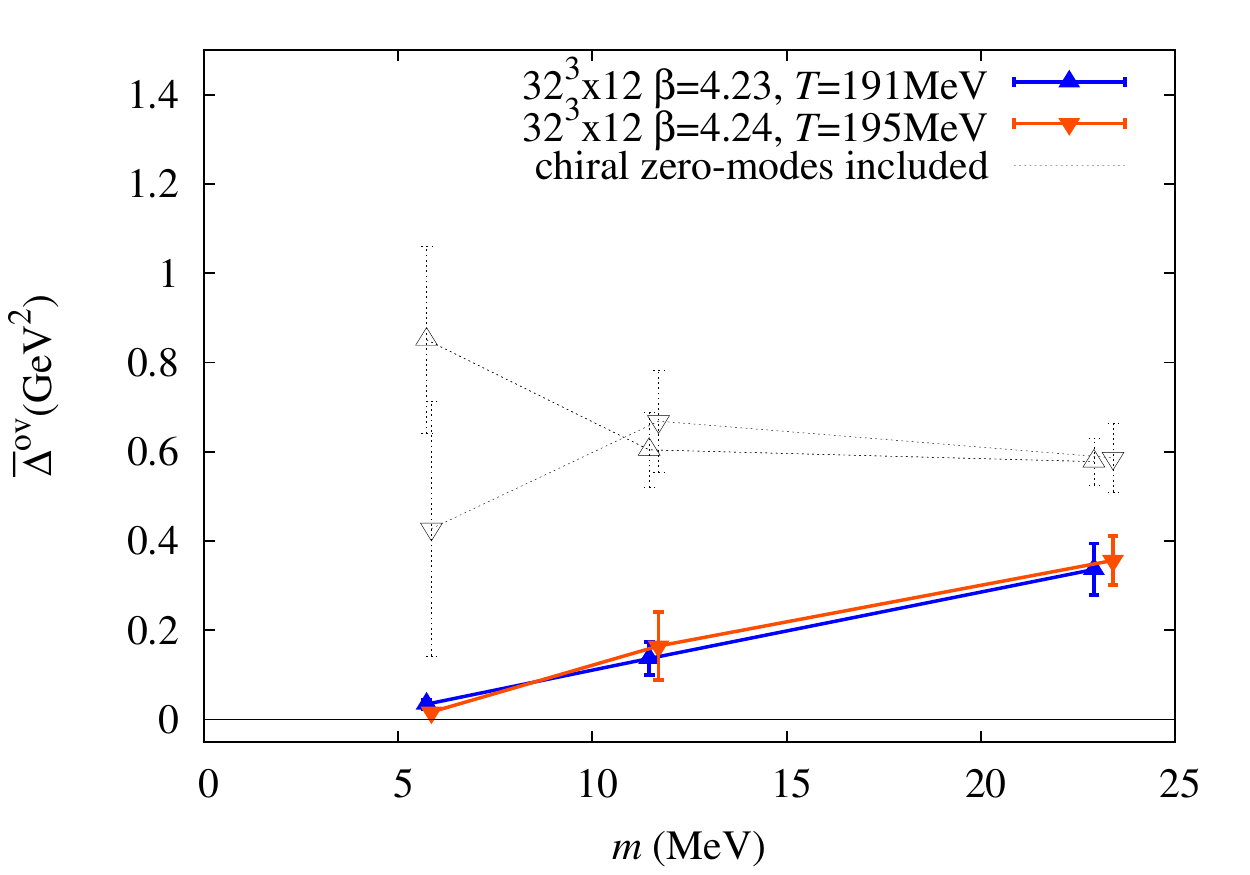}
  \caption{
  The quark mass dependence of $\bar{\Delta}_{\pi-\delta}^{\rm ov}$ (solid symbols) and $\Delta_{\pi-\delta}^{\rm ov}$ (dashed).
  Data for coarse (left panel) and fine (right) lattices are shown \cite{Tomiya:2016jwr}.
  }
  \label{fig:figure/FinalAfterRW}
\end{figure*}

In Fig.~\ref{fig:figure/FinalAfterRW}, our results for $\bar{\Delta}_{\pi-\delta}^{\rm ov}$
(solid symbols) and $\Delta_{\pi-\delta}^{\rm ov}$ (dashed) are plotted.
We confirm that our data for $\bar{\Delta}_{\pi-\delta}^{\rm ov}$ are stable against 
the change of the lattice size and lattice spacing.
The chiral limit of $\bar{\Delta}_{\pi-\delta}^{\rm ov}$ is consistent with zero.
Precisely speaking, all our data are well
described (with $\chi^2/$d.o.f. $\lesssim 1$) by a simple linear function of $m$,
which becomes consistent with zero “before” the chiral limit.
We observe neither strong volume dependence nor $\beta$ dependence of this behavior.

Our preliminary data on finer lattices show a consistent result,
which is shown in a separate contribution in these proceedings \cite{Suzuki:lat17}.

\section{Topological susceptibility}

In our zero-temperature runs \cite{Aoki:2017paw},
the gluonic measurement of the topological susceptibility $\chi_t$
on configurations generated by the M\"obius domain-wall fermion
shows a good chiral behavior,
which is consistent with the chiral perturbation theory prediction.
We observe no significant dependence on the lattice spacing.

\begin{figure*}[tbp]
    \centering
    \includegraphics[width=8.0cm]{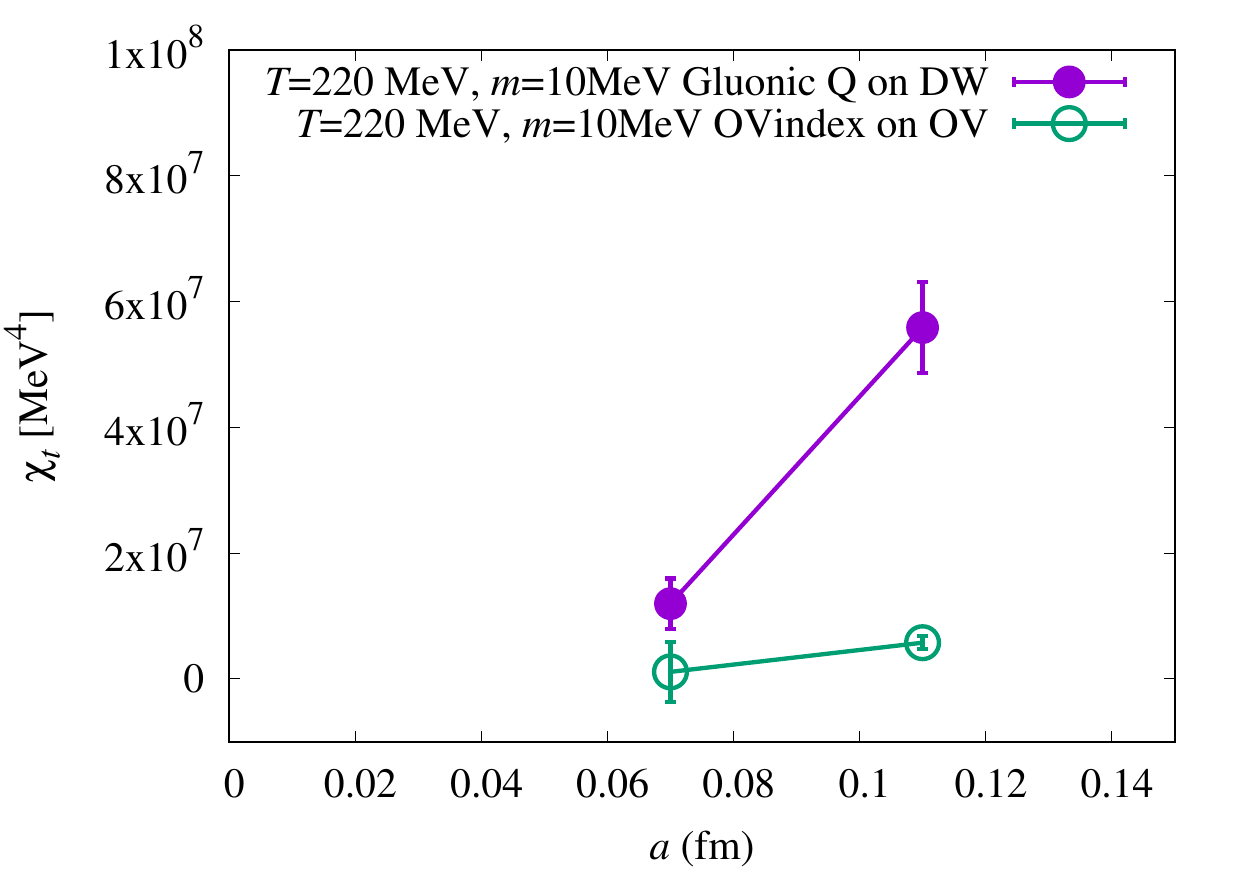}
  \caption{
    Dependence on the lattice spacing $a$ of the topological susceptibility,
    computed with gluonic definition with the M\"obius domain-wall sea quarks (filled symbols),
    and that with the index of the overlap Dirac operator with overlap/domain-wall reweighting (open).
    Data at $T\sim 220$ MeV are presented.
  }
  \label{fig:chit-a}
\end{figure*}

At high temperature, we, however, see a strong cut-off dependence
of $\chi_t$ as presented by the filled symbols
in Fig.~\ref{fig:chit-a}, which are obtained with
a gluonic definition of the topological charge 
after the Wilson flow $\sqrt{8t} \sim 0.5$fm.
It implies that the overlap/domain-wall reweighting is essential when
one simulates high temperature QCD with $a>0.1$ fm.
Once it is performed, we observe
a milder $a$ dependence (open symbols).
In the following, we mainly report on our preliminary results
on the finest lattice at $a\sim 0.07$fm,
where the M\"obius domain-wall fermion and the reweighted overlap fermion
give a consistent result within the statistical accuracy.
For the overlap case, we use its index as the definition of the topological charge.
See \cite{Yasumichi:lat17} for more details. 

\begin{figure*}[tbp]
    \centering
    \includegraphics[width=8.0cm]{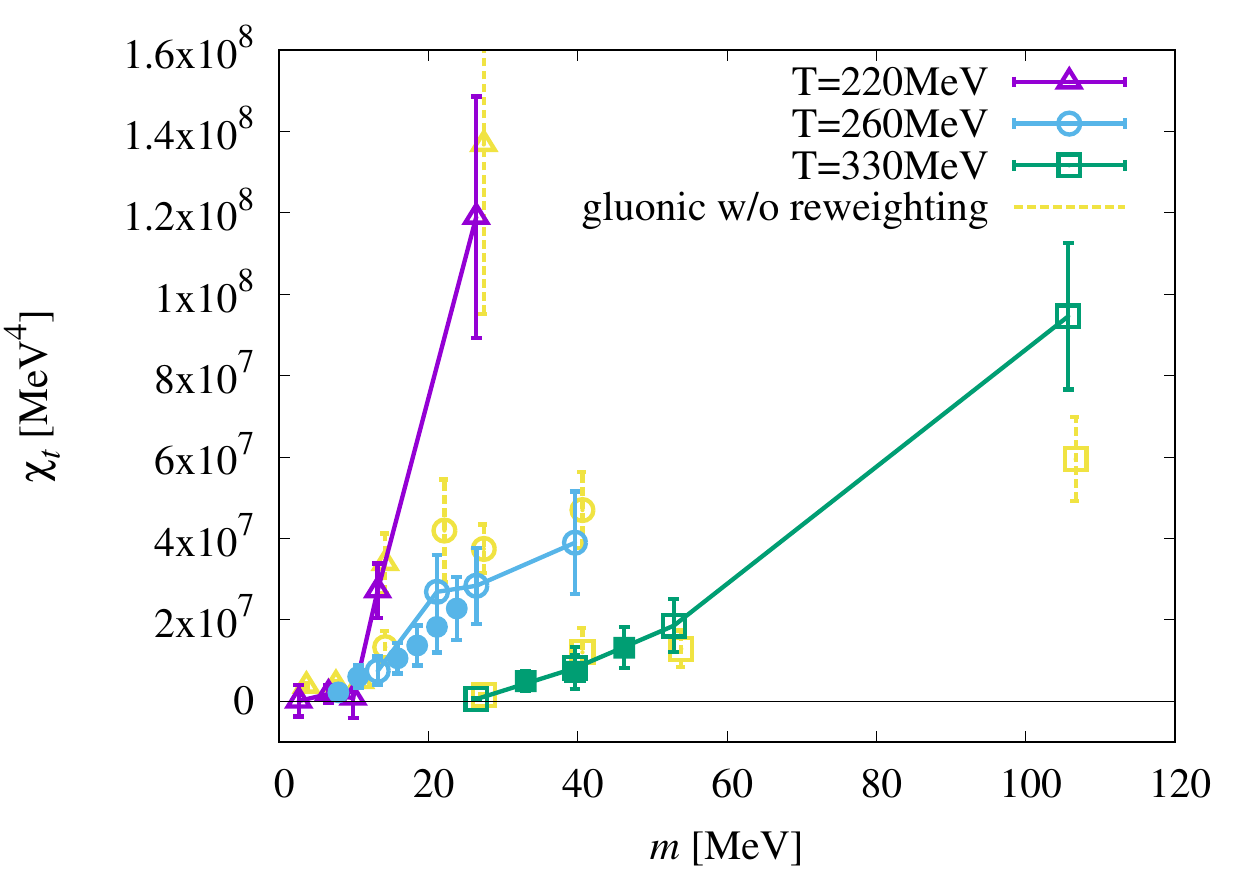}
    \caption{
      Preliminary result for $\chi_t$ at $\beta=4.30$ at $L_t=12,10,8$, which corresponds to
      $T=$220 (triangle symbols), 260(circles), 330(squares) MeV, respectively.
      The solid symbols are obtained from the index of the overlap Dirac operator with the reweighting,
      while the dashed symbols are those from the gluonic definition on the original configurations with M\"obius domain-wall fermions.
  }
  \label{fig:chit-summary}
\end{figure*}

\begin{figure}[tbp]
   \centering
   \subfigure[$\chi_t$ at $T=220$ MeV with two different lattice volumes.
             ]%
             {\includegraphics[width=0.475\textwidth]{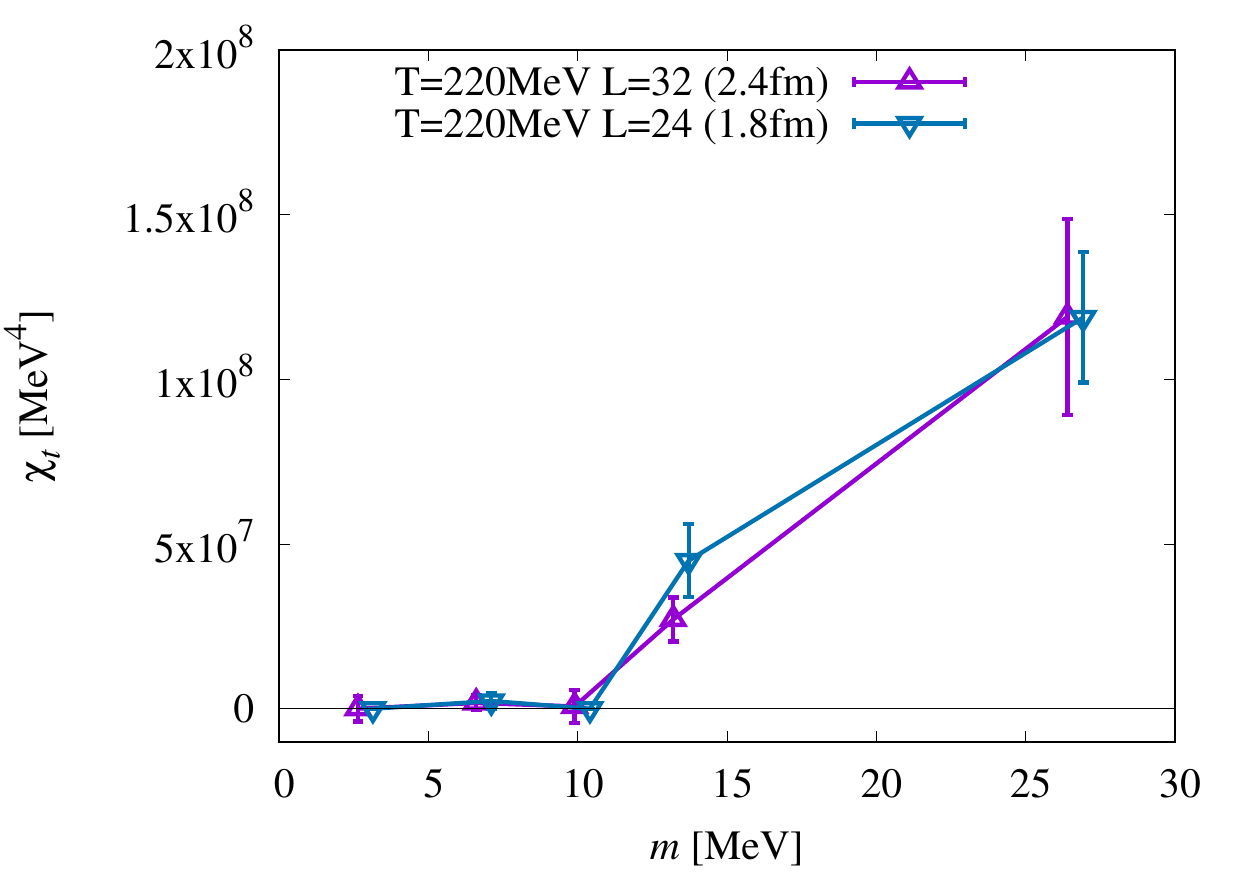}}\hfill
             \subfigure[$\chi_t$ at $T=330$ MeV with two different lattice volumes.
             ]%
             {\includegraphics[width=0.475\textwidth]{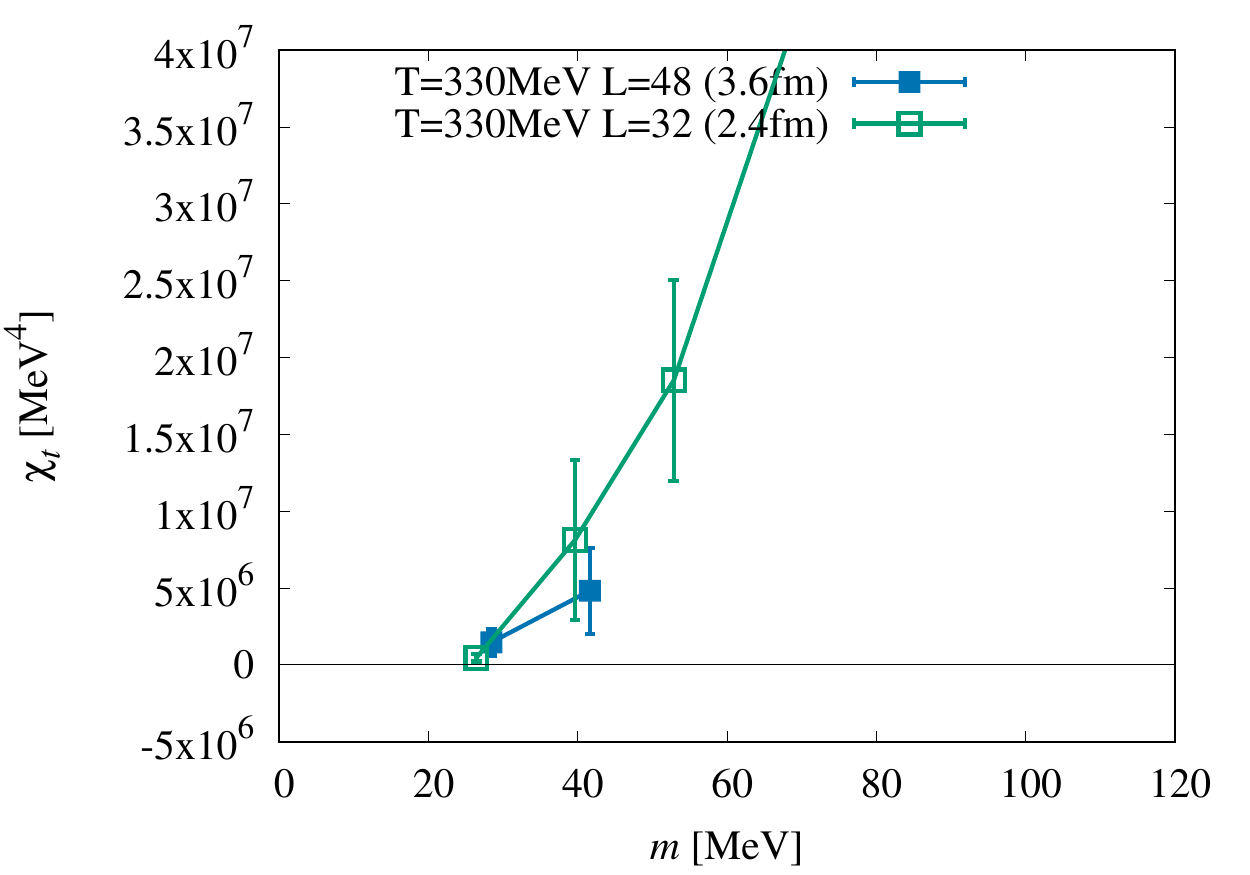}}\hfill
   \subfigure[$\chi_t$ at $T=220$ MeV with two different lattice spacings.
             ]%
             {\includegraphics[width=0.475\textwidth]{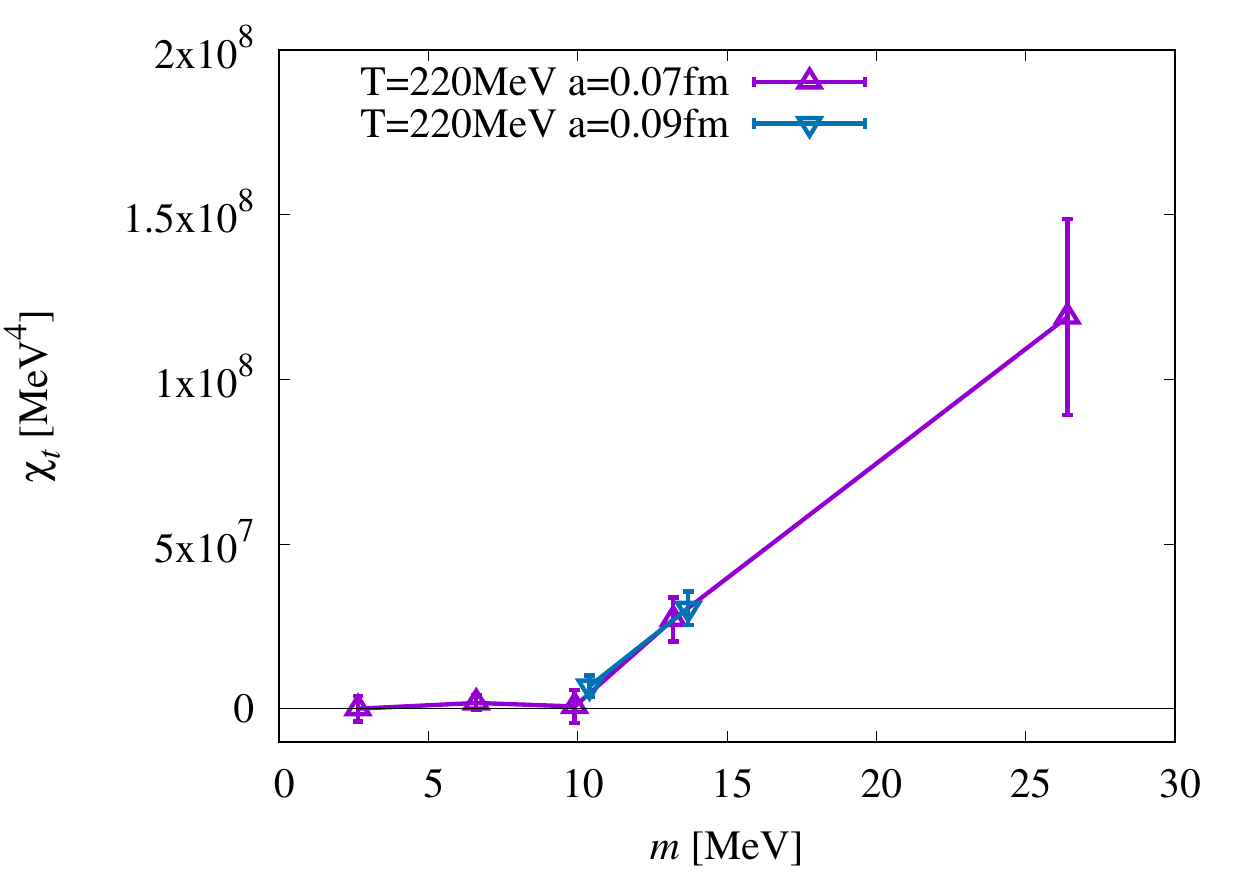}}\hfill
   \caption{Comparison of topological susceptibility with different volumes and different lattice spacings (preliminary).}
   \label{fig:finiteVfinitea}
\end{figure}

Figure~\ref{fig:chit-summary} summarizes our preliminary results
at our finest lattice spacing $a\sim 0.07$ fm at three different
temporal sizes $L_t=12,10,8$, which corresponds to
$T=$220 (triangle symbols), 260 (circles), 330 (squares) MeV, respectively.
The solid symbols are obtained from the index of the overlap Dirac operator with the reweighting,
which are consistent with the dashed symbols obtained
from the gluonic definition on the original configurations with M\"obius domain-wall fermions.
The data at three different temperatures show a sharp drop of the topological susceptibility
near the chiral limit but at slightly different values of the quark mass,
which is not observed by other groups \cite{Bonati:2015vqz, Petreczky:2016vrs, Borsanyi:2016ksw}.
They are consistent with zero
at small but a finite value of the quark mass, which is consistent with the existence of
the critical mass $m_{cr}$ predicted in (\ref{eq:mcritical}).
Our data at different volumes and different lattice spacings are
consistent, as shown in Fig.~\ref{fig:finiteVfinitea}.

Here let us discuss a naive dimensional analysis of $\chi_t$.
At zero temperature, it is inferred from the chiral effective theory that
\begin{eqnarray}
  \chi_t = \frac{|m|\Sigma}{N_f}.
\end{eqnarray}
The discontinuity of its $m$ derivative is the
sign of SSB of the $SU(2)_L\times SU(2)_R$ symmetry.
Above the critical temperature, $\chi_t$ must have
even power of $m$ near the chiral limit,
and should start with the quadratic term
\begin{eqnarray}
  \chi_t = m^2\mu^2 +\cdots,
\end{eqnarray}
where $\mu$ is some unknown scale given by the QCD dynamics
at finite temperature $T$.
A simple choice $\mu = T$ is not allowed since
we know that $\chi_t$ vanishes at $T= \infty$.
The dilute instanton gas (DIGA) model \cite{Gross:1980br} suggests
\begin{equation}
  \label{eq:DIGA}
  \mu^2 = \Lambda_{QCD}^2  (\Lambda_{QCD}/T)^\alpha,
\end{equation}
with a positive power $\alpha$.
This structure poses a ``naturalness'' question of
a generation of the very small $\mu=\Lambda_{QCD} (\Lambda_{QCD}/T)^{\alpha/2}$ originated
from the very high energy $T$.
If (\ref{eq:DIGA}) is the case, it requires a long-range correlation
at very high temperature, which is questionable
because it implies
a long-range mode in quark-gluon plasma.
Our data suggest  $\chi_t=0$ at finite quark mass,
as an alternative answer to this naive dimensional puzzle.
We do not need the small scale $\mu$ to be generated.

If the existence of the critical
quark mass $m_{cr}$ below which $\chi_t=0$ is confirmed,
the topological susceptibility plays a role of an order parameter
of the QCD phase transition even at finite quark mass.
Then, the phase transition is likely to be the first order,
since there is no symmetry enhancement produce a long-distance correlation
at finite quark mass.

Another important phenomenological issue is the
impact on the axion dark-matter scenario \cite{Moore:2017ond}.
The axion mass squared $m_a^2$ at finite temperature is proportional to $\chi_t$.
Therefore, if $\chi_t=0$, the axion mass is zero,
which is inconsistent with the current cosmological bound,
since zero (or too small) axion mass means
too much dark matters produced in the early universe
to keep the current size of the universe.
Our data do not exclude tiny non-zero
value of the topological susceptibility
but the sudden drop at finite quark mass is informative.
We also note that the strange quark is quenched in our simulations,
which may lead to a different phase from the nature.

\section{Summary and discussion}

We have discussed a possible disappearance of the axial $U(1)$ anomaly
above the critical temperature of the QCD phase transition.

First, we have shown that the SSB of $SU(2)_L\times SU(2)_R$ and
the axial $U(1)$ anomaly are not independent, but tightly connected
to each other, at least.
This can be confirmed by decomposing the observables into eigenmodes of the Dirac operator.

Next, we have explained our numerical set-up of QCD simulation
employing chiral fermions.
We have found an exceptionally large violation of the Ginsparg-Wilson relation
in the low-lying eigenmodes of the M\"obius domain-wall Dirac operator,
and the $U(1)$ anomaly is sensitive to these bad modes.
Therefore, we employed the overlap fermion for the observables,
together with the reweighting of the fermion determinant.
This overlap/domain-wall reweighting is essential for $a> 0.1$ fm,
since the partially quenching artifact is enhanced by the bad modes.

Our data for the $U(1)_A$ susceptibility is
consistent with zero near the chiral limit at temperatures
in the region 190--339 MeV.
The quark mass dependence of the topological susceptibility 
shows a sharp drop, which is consistent with
the existence of the critical mass $m_{cr}$ below which
$\chi_t$ is zero.

If our data really indicate the absence of the
$U(1)$ anomaly above the critical temperature
(for thermo-dynamical quantities, at least),
they give phenomenological impacts on
the QCD phase diagram, axion dark matter scenario, and so on.

As a final remark, we comment on the ``U(1) anomaly session'',
where this presentation was given.
In this session, three talks on lattice QCD simulations
gave some different results,
but they all showed a difficulty
of finite temperature QCD, due to growth of the lattice artifacts.
In particular, it is difficult to precisely estimate the
topological susceptibility at temperatures higher than 300 MeV,
which is necessary for estimating the axion mass produced in the early universe.
We need more studies of the systematics and a careful continuum limit.

I thank L. Glozman, K. Hashimoto, Y. Iwasaki, K. Kanaya,  T. Kanazawa, S. Prelovsek, C. Rohrhofer, 
and Y. Taniguchi for useful discussions.
I also thank the members of the JLQCD collaboration for their supports.
Numerical simulations are performed on IBM System Blue Gene Solution at KEK under
a support of its Large Scale Simulation Program (No. 16/17-14),
and on Oakforest-PACS supercomputer at the CCS, University of Tsukuba.
This research also used computational resources of the HPCI system provided by
Joint Center for Advanced High Performance Computing (JCAHPC)
through the HPCI System Research Projects (Project ID: hp170061).
This work is supported in part by the Japanese Grant-in-Aid for Scientific Research (No. JP26247043)
and by MEXT as “Priority Issue on Post-K computer”
(Elucidation of the Fundamental Laws and Evolution of the Universe)
and by Joint Institute for Computational Fundamental Science (JICFuS).

\end{document}